\documentstyle[preprint,aps]{revtex}

\begin{document}

\draft

\title{Quantized Fields and Temperature in Charged Dilatonic Black
Hole Spacetimes}

\author{Daniel J.  Loranz\cite{Lor} and William A.\ Hiscock\cite{His}}

\address{Department of Physics, Montana State University, Bozeman,
Montana 59717}

\date{July 22, 1996}

\maketitle

\begin{abstract} The stress-energy tensor of a quantized scalar field
is computed in the reduced two-dimensional charged dilatonic black
hole spacetime of Garfinkle, Horowitz, and Strominger.  In order for
the stress-energy of quantized fields to be regular on the event
horizon in both the extreme string metric and the conformally
associated physical metric, it is necessary to assign a nonzero
temperature, $T = (8\pi e^{\phi_0} M)^{-1}$, to the extreme string
metric, contrary to the expectation that this horizonless spacetime
would have a natural temperature of zero.

\end{abstract}

\pacs{ }

\section{}

Garfinkle, Horowitz, and Strominger\cite{GHS}(hereafter, GHS) found
static spherical charged black hole solutions in the low-energy
approximation to string theory.  These solutions possess nonconstant
dilaton fields in addition to the electromagnetic field $F_{\mu\nu}$.
The GHS black hole solutions have substantially different properties
compared to the analogous Reissner-Nordstr\"{o}m black hole solutions
in general relativity.  In particular, the GHS black hole lacks inner
horizons, and while a maximal value of the charge exists which, in
both sets of solutions, separates the black hole solutions from naked
singularities, these extreme solutions have quite different properties
in the two theories.  The event horizon of the extreme GHS solution in
what we call the physical metric, $g_{\mu\nu}$, is singular, whereas
the extreme Reissner-Nordstr\"{o}m black hole has a nonsingular event
horizon with well defined stress-energy\cite{AHL}.  However, as noted
by Garfinkle, Horowitz, and Strominger, the singular nature of the
event horizon in the physical metric is irrelevant.  Strings do not
couple to the physical metric $g_{\mu\nu}$ but to the conformally
related metric $e^{2\phi}g_{\mu\nu}$, which we will call the string
metric, where $\phi$ is the dilaton field.  In fact, in the string GHS
metric, the extreme spacetime has no event horizon and is geodesically
complete.

The thermodynamics of the Reissner-Nordstr\"{o}m and GHS spacetimes
also differs.  In the Reissner-Nordstr\"{o}m spacetime the black hole
temperature decreases steadily from the Schwarzschild value to zero as
the extreme limit is approached.  In comparison the string GHS metric
has a nonzero temperature independent of the charge, until the extreme
solution is reached.  Then the temperature apparently shifts
discontinuously to zero since the extreme spacetime has no horizon.

In this paper, working in two-dimensional black hole metrics obtained
by discarding the angular portions of the GHS metrics, we calculate
the expectation value of the stress-energy tensor for a quantized
conformally coupled massless scalar field in the Hartle-Hawking state.
We find that by assigning the extreme string spacetime of GHS a
nonzero temperature equal to that of the nonextreme string spacetime,
it is possible to render $\langle T_{\mu\nu} \rangle$ regular
everywhere outside and on the horizon for both the string and physical
metrics.

The effective Lagrangian for low-energy string theory used by
Garfinkle, Horowitz, and Strominger is
\begin{equation}
	L = \int d^4x {\sqrt {-g}}[-R+2(\nabla\phi)^2
	+e^{-2\phi}F^2 ]  ,
	\label{Lagrangian}
\end{equation}

where $\phi$ is the dilaton field and $F_{\mu\nu}$ is the Maxwell
field.  Static, spherically symmetric black hole solutions to the
field equations obtained from Eq.(\ref{Lagrangian}) are described
by\cite{GHS}:

\begin{equation}
	ds^2 = - \left( 1 - {{2M} \over r} \right) dt^2 + \left( 1
	- {{2M}	\over r} \right)^{-1} dr^2 + r \left(r - {{Q^2 
	e^{-2\phi_0}} \over M} \right) d\Omega^2   ,
	\label{physmetric}
\end{equation}
\begin{equation}
	e^{-2\phi} = e^{-2\phi_0} \left( 1 - {{Q^2 e^{-2\phi_0}}
	\over {Mr}} \right)   ,
	\label{dilaton}
\end{equation}
and
\begin{equation}
	F = Q \sin \theta d \theta \wedge d \phi .
	\label{maxwell}
\end{equation}
Here $d\Omega^2$ is the metric of the two-sphere, $\phi_0$ is the
asymptotic value of the dilaton field, and we have corrected the
errors in the powers of $e^{\phi_0}$ in accordance with the Erratum to
Ref.\cite{GHS}.  The surface
\begin{equation}
	r = {{Q^2 e^{-2\phi_0}} \over M}
	\label{singularity}
\end{equation}

is singular, and the extreme limit occurs when the charge is increased
to a value sufficient to bring this surface into coincidence with the
horizon at $r = 2M$, namely when
\begin{equation}
	Q^2 = 2 M^2 e^{2\phi_0} .
	\label{extreme}
\end{equation}
The strings do not couple to the physical metric $g_{\mu\nu}$ of
Eq.(\ref{physmetric}) but rather to the conformally related string
metric $e^{2\phi}g_{\mu\nu}$.  Applying this conformal transformation
to Eq.(\ref{physmetric}) yields
\begin{equation}
	d{s_{string}}^2 = - { {\left( 1-{{2M e^{\phi_0}}\over \rho}
	\right)}\over {\left( 1 - {{Q^2 e^{-\phi_0}} \over {M \rho}}
	\right)} } d\tau^2 + {{d\rho^2} \over {\left(1 - 
	{{2Me^{\phi_0}}\over \rho} \right)\left( 1 - {{Q^2 
	e^{-\phi_0}} \over {M\rho}} \right) }} + \rho^2
	d\Omega^2   ,
	\label{stringmetric}
\end{equation}
where the coordinates $(\tau,\rho)$ are defined by $\tau = e^{\phi_0}
t$, $\rho = e^{\phi_0} r$.  In the extreme limit, when $Q^2 = 2 M^2
e^{\phi_0}$, Eq.(\ref{stringmetric}) reduces to
\begin{equation}
	ds^2 = - d\tau^2 + \left( 1 - {{2Me^{\phi_0}} \over \rho}
	\right)^{-2} d\rho^2 + \rho^2 d\Omega^2  .
	\label{extremestring}
\end{equation}
This extreme geometry has no horizon, and the spatial geometry of
constant $\tau$ hypersurfaces is geometrically identical to that of
$t$ = constant surfaces in the extreme Reissner-Nordstr\"{o}m metric.
As a result, the $\rho = 2e^{\phi_0}M$ surface is at an infinite
proper distance from any point in the manifold with $\rho >
2e^{\phi_0}M$.  In the GHS extreme spacetime, the distance is infinite
in any direction, whereas in the extreme Reissner-Nordstr\"{o}m case,
the distance to $r = M$ is infinite only in spacelike directions.

Since the GHS metric is the low-energy approximation to full string
theory black hole solutions, it is sensible to examine the physics of
ordinary quantized free fields (as opposed to quantized strings) in
the GHS black hole background.  While obtaining values for tensor
objects such as $\langle \psi^2 \rangle$ or $\langle T_{\mu\nu}
\rangle$ for a quantized field $\psi$ can be quite difficult in curved
four-dimensional spacetimes, calculating $\langle T_{\mu\nu} \rangle$
for a conformally coupled field in a two-dimensional spacetime is both
a straightforward and often valuable exercise\cite{DFU,WH1,LHA}.  In
this paper we restrict our attention to two dimensions, computing
$\langle T_{\mu\nu} \rangle$ for a conformally coupled quantized
scalar field in the two-dimensional metrics obtained from
Eqs.(\ref{physmetric},\ref {stringmetric}) by setting the angular
coordinates $\theta$ and $\phi$ to constant values.  We examine the
behavior of a quantized field in both the string and physical metrics.

The stress-energy tensor of a quantized scalar field in a
two-dimensional black hole spacetime for the Hartle-Hawking state may
be easily calculated; for convenience, in this paper we will use the
approach of Ref.\cite{LHA}.  The spatial coordinate of the black hole
metric must first be transformed in order to place the metric in
``Schwarzschild gauge''
\begin{equation} 
	ds^2 = - f(r)dt^2 + {{dr^2} \over {f(r)}}.
	\label{Schgauge}
\end{equation}
The temperature defined by the geometry is then given by 
\begin{equation}
	T = {{f'|_{r_0}} \over {4 \pi}} ,
	\label{tempdef} 
\end{equation} 
where $r_0$ is the radius of the event horizon and a prime denotes
differentiation with respect to $r$.  The trace of the stress-energy
tensor is given by the conformal anomaly,
\begin{equation}
	 \langle {T_\alpha}^\alpha \rangle = - {f'' \over {24 \pi}}.
	 \label{tracegeneral} 
\end{equation}
The components may then be completely determined by quadrature of the
conservation equation, with regularity of the stress-energy tensor on
the black hole horizon fixing the integration constant.  Specifically,
\begin{equation} 
	\langle {T_r}^r \rangle = -{{{f'}^2} \over {96 \pi f}} 
	+ {{\pi T^2} \over {6 f}} , 
	\label{Trrgen}
\end{equation} 
\begin{equation} 
	\langle {T_t}^t \rangle = \langle{T_\alpha}^\alpha \rangle 
	- \langle {T_r}^r \rangle .  
	\label{Tttgen}
\end{equation}

The physical GHS metric of Eq.(\ref{physmetric}), when restricted to
two dimensions, becomes simply the two-dimensional Schwarzschild
metric for all values of $Q$.  No transformation is needed and one can
quickly calculate that the spacetime has a temperature of
\begin{equation}
	T_{Sch} = {1 \over {8\pi M}} ,
	\label{Tsch}
\end{equation}
while the components of the stress-energy tensor for the quantized
field are found to be
\begin{equation}
	\langle {T_t}^t \rangle = {{56M^3-4M^2r-2Mr^2-r^3} \over
		{384\pi M^2 r^3}} ,
	\label{Tttsch}
\end{equation}
\begin{equation}
	\langle {T_r}^r \rangle = {1 \over {384 \pi M^2}}
	\left( 1 + {{2M}\over r}\right)\left(1 + {{4M^2}
	\over {r^2}}\right) ,
	\label{Trrsch}
\end{equation}
and the trace anomaly is
\begin{equation}
	\langle {T_\alpha}^\alpha \rangle = {M \over {6 \pi r^3}} .
	\label{tracesch}
\end{equation}

In order to calculate the stress-energy tensor of a quantized field in
the GHS string metric described by Eq.(\ref{stringmetric}) restricted
to two dimensions, the metric must first be transformed to
Schwarzschild gauge, which is accomplished by defining a new spatial
coordinate $x$,
\begin{equation}
	x = \int \left( 1 - {{e^{-\phi_0}Q^2} \over {M \rho}} 
	\right)^{-1} d\rho .
	\label{stringschwcoord}
\end{equation}
The derivatives in Eqs.(\ref{tempdef}-\ref{Tttgen}) are now taken with
respect to the new coordinate $x$.  We shall, however, express the
resulting stress-energy tensor components in terms of the original
$(\tau,\rho)$ coordinates of Eq.(\ref{stringmetric}).  For all
nonextreme values of $Q$, the string metric has temperature
\begin{equation}
	T_{string}  = {1 \over {8 \pi e^{\phi_0} M}} .
	\label{Tstring}
\end{equation}
To obtain the components of the stress-energy tensor one may either
directly integrate the conservation equation as we did for the
physical metric, or utilize the relation between stress-energy tensors
of conformally invariant fields in conformally related
spacetimes\cite{BD}.  In two dimensions, if two geometries are related
by a conformal factor $\Omega(x)$, such that ${\bar g}_{\mu\nu} =
\Omega^2(x) g_{\mu\nu}$, then the vacuum stress-energy tensors of a
conformally invariant scalar or spinor field are related by:
\begin{eqnarray}
	\langle {{\bar T}_\mu}^\nu \rangle_{ren} = & \left({g 
	\over {\bar g}}	\right)^{1/2}\langle {T_\mu}^\nu 
	\rangle_{ren} + \left({1 \over {12\pi}}\right)
	\Big[\left(\Omega^{-3} \Omega_{;\alpha\mu} -2\Omega^{-4}
	\Omega_{,\alpha}\Omega_{,\mu}\right)g^{\alpha\nu} 
	 \nonumber \\ & +{\delta_\mu}^\nu g^{\alpha\sigma}
	 \left({3 \over 2} \Omega^{-4} \Omega_{,\alpha}
	 \Omega_{,\sigma}-\Omega^{-3} \Omega_{;\alpha\sigma} 
	 \right)\Big],
       \label{confrel}
\end{eqnarray}
where all derivatives are taken with respect to the unbarred metric.
The components of the stress-energy tensor for the quantized scalar
field in the string spacetime are found to be
\begin{eqnarray}        
	\langle {T_r}^r \rangle =&\left[384\pi e^{3\phi_0}M^3\rho^2
		(e^{\phi_0}M\rho - Q^2)\right]^{-1}\Big[8 
		e^{5\phi_0} M^5 - 8e^{3\phi_0} M^3 Q^2 + 2 
		e^{\phi_0} M Q^4 \nonumber \\ & +(4 e^{4\phi_0} M^4
		- 4 e^{2\phi_0}M^2 Q^2 + Q^4) \rho+(2e^{3\phi_0}M^3
		-2e^{\phi_0}MQ^2)\rho^2	+e^{2\phi_0}M^2\rho^3\Big] ,
	\label{Trrstring}
\end{eqnarray}
\begin{eqnarray}
	\langle {T_\tau}^\tau \rangle = &\left[384\pi e^{3\phi_0}
	M^3\rho^3(e^{\phi_0}M\rho-Q^2)\right]^{-1}
	\Big[-32e^{4\phi_0}M^4Q^2+16e^{2\phi_0}M^2 Q^4 \nonumber 
	\\ & +(56e^{5\phi_0}M^5	-24e^{3\phi_0}M^3Q^2-2e^{\phi_0}
	MQ^4)\rho -(4e^{4\phi_0}M^4-4e^{2\phi_0}M^2Q^2+Q^4)
	\rho^2 \nonumber \\ & -(2e^{3\phi_0}M^3-2e^{\phi_0}MQ^2)
	\rho^3 -e^{2\phi_0}M^2\rho^4\Big]  ,
	\label{Tttstring}
\end{eqnarray}
and the trace anomaly is given by
\begin{equation}
	\langle {T_\alpha}^\alpha \rangle = {{(Q^2-e^{2\phi_0}
	M^2)(Q^2 - 2 e^{\phi_0}M\rho)} \over {24 \pi e^{\phi_0}
	M\rho^3(e^{\phi_0}M\rho-Q^2)}}  .
	\label{tracestring}
\end{equation}

It is interesting to compare the physical behavior of the
stress-energy of the quantized field in the GHS charged black hole
string metric, given by Eqs.(\ref{Trrstring}-\ref{tracestring}) with
those computed for the usual Reissner-Nordstr\"{o}m black hole (these
may be obtained from the expressions for the Unruh vacuum components
given in Ref.\cite{WH1} by a simple transformation).  For the GHS
black hole, the energy density as seen by a static observer $\epsilon
= -\langle{T_\tau}^\tau\rangle$ is always negative in the neighborhood
of the horizon; the radial stress is similarly always positive,
indicating a pressure.  In comparison, the energy density of a
quantized field near the horizon of a two-dimensional
Reissner-Nordstr\"{o}m black hole is negative for $Q^2 < 8M^2/9$, but
is positive for larger values of $Q^2$.  The radial stress is also
positive for $Q^2 < 8M^2/9$, and negative for larger values of $Q^2$.

The case of the extreme GHS black hole, with $Q^2 = 2e^{2\phi_0}M^2$,
must be treated separately, as it possesses no horizon.  In fact, once
the metric of Eq.(\ref{extremestring}) is reduced to two dimensional
form, it is flat, and hence has no trace anomaly.  The lack of a
horizon means no geometrically defined temperature exists in this
case.  Garfinkle, Horowitz, and Strominger suggest zero temperature as
the natural state for this spacetime, even though the temperature is
then a discontinuous function of $Q^2$.  Alternatively, one could
choose the temperature to be continuous by assigning a value
$T_{extreme} = (8\pi e^{\phi_0} M)^{-1}$.  If an arbitrary temperature
$T_{extreme}$ is temporarily assigned, then the most general
time-reveral symmetric ({\it i.e.}, equilibrium) solution of the
stress-energy conservation equation is that of a simple boson gas:
\begin{equation}
	\langle {T_\rho}^\rho \rangle = - \langle {T_\tau}^\tau
	\rangle	= {{\pi} \over {6}}T_{extreme}^2  .
	\label{extremetmn}
\end{equation}

Consider now the temperatures associated with both the physical and
string metrics.  The physical metric is always precisely Schwarzschild
(independent of $Q$) in two dimensions, and hence must always have $T
= (8\pi M)^{-1}$; assignment of any other temperature would lead to a
divergent stress-energy for the quantized field on the
horizon\cite{LHA}.  Similarly, the string metric's natural temperature
is $T= (8\pi e^{\phi_0} M)^{-1}$ for all values $Q^2 < e^{2\phi_0} M^2
$.  Again, assignment of any other temperature would cause a strong
divergence in the stress-energy on the horizon.  These natural,
geometrically defined temperatures and the associated stress-energy
tensors for quantized fields are also related directly through the
conformal transformation of Eq.(\ref{confrel}).

In the extreme case, the physical metric is still precisely
Schwarzschild, and thus the spacetime must either be assigned the
usual temperature or suffer divergent stress-energy on the horizon.
In the extreme string metric, though, there is no horizon, and hence
any temperature (including zero) can now be assigned to the spacetime
without causing singular behavior in the quantized fields in this
metric.  However, the conformal relation of Eq.(\ref{confrel}) picks
out a unique temperature for the extreme string metric.  If the
quantized fields are to be regular in both the physical and string
metrics, one must choose $ T_{extreme} = (8\pi e^{\phi_0}M)^{-1} $.
Any other choice in the string metric will lead to divergences on the
horizon of the physical metric by Eq.(\ref{confrel}).  Thus, it
appears that the natural choice for the temperature of the extreme
string metric is not zero, despite the lack of horizon, but rather $
T_{extreme} = (8\pi e^{\phi_0}M)^{-1} $.  This choice preserves
continuity of the temperature in the string metric, and is the only
choice which will allow quantized fields in both the string and
physical metrics to be regular outside and on all horizons.

It is interesting to compare and contrast this result to the case of
the extreme Reissner-Nordstr\"{o}m black hole \cite{AHL} in four
dimensions.  Hawking, Horowitz, and Ross \cite{HHR}noted that the
Euclidean section of the extreme Reissner-Nordstr\"{o}m metric allowed
one to identify the geometry with arbitrary period in Euclidean time
(and hence arbitrary temperature) without creating a conical
singularity in the Euclidean spacetime.  Loranz, Hiscock, and Anderson
\cite{LHA} demonstrated, however, that the stress-energy tensor of a
quantized field would diverge strongly on the event horizon in the
Lorentzian spacetime unless the temperature was chosen to be zero, the
value obtained by extrapolating the form of the temperature function
$T(M,Q)$ from the nonextreme Reissner-Nordstr\"{o}m black hole.  In
the present case, we again find that there will be a divergence of the
stress-energy of a quantized field (in this case, for a field which
takes values on the conformally related physical metric) if the
temperature of the extreme GHS string metric is taken to be a value
other than that extrapolated from the nonextreme GHS black hole
temperature function.  On the other hand, in this case, it means the
extreme GHS metric must be assigned a nonzero temperature, whereas in
the extreme Reissner-Nordstr\"{o}m case the natural temperature was
zero.

\acknowledgements 
The work of W.\ A.\ H.  was supported in part by
National Science Foundation Grant No.  PHY-9511794.

\end{document}